\documentclass[10pt,twoside,twocolumn,english,aps,manuscript,aps,preprint,show pacs,prb,superscriptaddress]{revtex4}
\usepackage[T1]{fontenc}
\usepackage[latin9]{inputenc}
\usepackage{babel}
\usepackage{textcomp}
\usepackage{amstext}
\usepackage{graphicx}
\usepackage{esint}
\usepackage{subscript}
\usepackage[unicode=true,pdfusetitle,
 bookmarks=true,bookmarksnumbered=false,bookmarksopen=false,
 breaklinks=false,pdfborder={0 0 1},backref=false,colorlinks=false]
 {hyperref}
\usepackage{breakurl}

\makeatletter

\DeclareRobustCommand{\greektext}{%
  \fontencoding{LGR}\selectfont\def\encodingdefault{LGR}}
\DeclareRobustCommand{\textgreek}[1]{\leavevmode{\greektext #1}}
\DeclareFontEncoding{LGR}{}{}
\DeclareTextSymbol{\~}{LGR}{126}
\newcommand{\lyxmathsym}[1]{\ifmmode\begingroup\def\b@ld{bold}
  \text{\ifx\math@version\b@ld\bfseries\fi#1}\endgroup\else#1\fi}

\providecommand{\tabularnewline}{\\}
\newcommand{\lyxdot}{.}

\@ifundefined{textcolor}{}
{%
 \definecolor{BLACK}{gray}{0}
 \definecolor{WHITE}{gray}{1}
 \definecolor{RED}{rgb}{1,0,0}
 \definecolor{GREEN}{rgb}{0,1,0}
 \definecolor{BLUE}{rgb}{0,0,1}
 \definecolor{CYAN}{cmyk}{1,0,0,0}
 \definecolor{MAGENTA}{cmyk}{0,1,0,0}
 \definecolor{YELLOW}{cmyk}{0,0,1,0}
}

\makeatother

\begin{document}

\title{BaV$_{\text{3}}$O$_{\text{8}}$: A possible Majumdar-Ghosh system
with $S$=1/2 }

\author{T. Chakrabarty}

\affiliation{Department of Physics, IIT Bombay, Powai, Mumbai 400076, India}

\author{A. V. Mahajan}

\email{mahajan@phy.iitb.ac.in}

\affiliation{Department of Physics, IIT Bombay, Powai, Mumbai 400076, India}

\author{A. A.Gippius}

\affiliation{Faculty of Physics, Moscow State University, Moscow 119991, Russia,
A.V. Shubnikov Institute of Crystallography, Moscow 119333, Russia}

\author{A. V. Tkachev}

\affiliation{Faculty of Physics, Moscow State University, Moscow 119991, Russia,
A.V. Shubnikov Institute of Crystallography, Moscow 119333, Russia}

\author{N. Büttgen}

\affiliation{Institute of Physics, University of Augsburg, D-86135 Augsburg, Germany}

\author{W. Kraetschmer}

\affiliation{Institute of Physics, University of Augsburg, D-86135 Augsburg, Germany}
\begin{abstract}
BaV$_{\text{3}}$O$_{\text{8}}$ contains magnetic V$^{\text{4+}}$($S=1/2$)
ions and also non-magnetic V$^{\text{5+}}$($S=0$) ions. The V$^{4+}$
ions are arranged in a coupled Majumdar-Ghosh chain like network.
A Curie-Weiss fit of our magnetic susceptibility $\chi(T)$ data in
the temperature region of $80-300$K yields a Curie constant $C=0.39$
cm$^{\text{3}}$K/mole V$^{\text{4+}}$ and an antiferromagnetic Weiss
temperature $\theta=-26\mathrm{K}$. The $\mbox{\ensuremath{\chi}}$(T)
curve shows a broad maximum at $T\simeq25$K indicative of short-range
order (SRO) and an anomaly corresponding to long-range order (LRO)
at $T_{\mathrm{N}}$$\sim6$K. The value of the \textquoteleft{}frustration
parameter\textquoteright{} ($f=|\theta/T_{\text{N}}|\sim5$) suggests
that the system is moderately frustrated. Above the LRO temperature
the experimental magnetic susceptibility data match well with the
coupled Majumdar-Ghosh (or $J_{nn}-J_{nnn}$ Heisenberg) chain model
with the ratio of the $nnn$ (next-nearest neighbor) to $nn$ (nearest
neighbor) magnetic coupling $\lyxmathsym{\textgreek{a}}$ = 2 and
$J_{nnn}$/$k_{B}$ = 40K. In a mean-field approach when considering
the inter-chain interactions, we obtain the total inter-chain coupling
to be about 16K. The LRO anomaly at $T_{\mathrm{N}}$ is also observed
in the specific heat $C_{\mathrm{P}}(T)$ data and is not sensitive
to an applied magnetic field up to $90$ kOe. A $^{\text{51}}$V NMR
signal corresponding to the non-magnetic vanadium was observed. Anomalies
at $6$K were observed in the variation with temperature of the $^{51}$V
NMR linewidth and the spin-lattice relaxation rate $1/T_{1}$ indicating
that they are sensitive to the LRO onset and fluctuations at the magnetic
V sites. The existence of two components (one short and another long)
is observed in the spin-spin relaxation rate $1/T_{2}$ data in the
vicinity of $T_{\mathrm{N}}$. The shorter component seems to be intimately
connected with the magnetically ordered state. We suggest that both
magnetically ordered and non-long range ordered (non-LRO) regions
coexist in this compound below the long range ordering temperature.
\end{abstract}


\maketitle

\section{introduction}

The field of low-dimensional and geometrically frustrated magnetism
is an active area of research in solid state physics.\cite{J.E.Greedan-Geometrically frustrated magnetic materials-2000,Physics Today-2006 A.P.Ramirez geometric frustration,G.ToulouseCmmun Phys-magnetic frustration}
In the last few decades, special emphasis has been laid on low-dimensional
spin systems\cite{SCience Anderson,Mermin Wagner} such as chains,
square lattices, ladders, especially after the discovery of high-temperature
superconductivity in cuprates.\cite{J.G. Bednorz HTSC-Ba-La-Cu-O-1986 -1}
In the case of one-dimensional (1D) antiferromagnetic (AF) chains,
the scenario becomes even more interesting if in addition to a nearest-neighbor
($nn$) interaction, a frustrating next-nearest-neighbor ($nnn$)
interaction is also present. Depending on the the ratio of the $nnn$
to $nn$ coupling ($J_{nnn}/J_{nn}$) in these so-called ``Majumdar-Ghosh''
(MG) chains (or $J_{nn}-J_{nnn}$ Heisenberg),\cite{Majumder-Ghosh  paper 1969}
distinct magnetic phases are formed. For $J_{nnn}/J_{nn}\geq0.24$,
the ground state is spontaneously dimerized with an energy gap in
the excitation spectrum.\cite{J2/J1>.24}

A number of Cu-based (3$d$$^{9}$) AF systems which can be described
by the MG (or $J_{nn}-J_{nnn}$ Heisenberg) chain model have been
investigated in the past. Some examples are CuCrO\textsubscript{4}\cite{CuCrO4},
(N\textsubscript{2}H\textsubscript{5})CuCl\textsubscript{3},\cite{(N2H5)CuCl3}
and (Cu(ampy)Br\textsubscript{2}).\cite{Cu(ampy)Br2 } Of these,
CuCrO\textsubscript{4} is thought to be close to the MG point($J_{nnn}/J_{nn}\simeq0.5$)\cite{CuCrO4}
and (N\textsubscript{2}H\textsubscript{5})CuCl\textsubscript{3}
is close to the quantum critical point ($J_{nnn}$/$J_{nn}$ =$0.2$4).\cite{(N2H5)CuCl3}
Surprisingly, it appears that the few vanadium-based (3$d$$^{1}$)
$S=1/2$ low-dimensional AF systems that have been investigated have
exclusively been either of the dimer type (CsV\textsubscript{2}O\textsubscript{5}\cite{CsV2O5Vspindimer}),
1D uniform chain type (NaV\textsubscript{2}O\textsubscript{5}\cite{NaV2O5})
or ladder type ((VO)\textsubscript{2}P\textsubscript{2}O\textsubscript{7}\cite{(VO)2P2O7},
CaV\textsubscript{2}O\textsubscript{5}\cite{CaV2O5}). On the other
hand, we were unable to find in literature any examples of vanadium-based
(3$d$$^{1}$) $S=1/2$ systems described by the $J_{nn}-J_{nnn}$
Heisenberg chain model. We have been exploring low-dimensional oxides
with the intention of unraveling novel magnetic properties. It seems
interesting to investigate systems where the magnetic exchanges arise
from the overlap of the (say) $d_{xy}$ orbitals (via the oxygen $p$)
rather than the $d_{x^{2}-y^{2}}$ orbitals as in the Cu-based systems.
In this paper, we report our studies on the yet unexplored system
BaV$_{\text{3}}$O$_{\text{8}}$ via magnetization, heat capacity
and $^{51}$V nuclear magnetic resonance (NMR) measurements. The susceptibility
$\chi(T)$ data exhibit a broad maximum at $25$K signifying short-range
order (SRO). The appearance of LRO at $T_{\mathrm{N}}$$\sim6$K is
evidenced in the susceptibility as well as the heat capacity data.
From our $\chi(T)$ data, we infer that the linkages between the magnetic
($S=1/2$) V$^{\text{4+}}$ ions in this compound are like those of
an MG chain with $\alpha\approx2$. The NMR signal from the magnetic
$^{51}$V ions could not be observed most probably due to the strong
on-site fluctuations of the V$^{4+}$ moment giving rise to a wipeout
of the NMR signal. On the other hand, a $^{51}$V NMR signal corresponding
to the nonmagnetic V$^{5+}$ nuclei was easily observed. Further,
the evolution of its lineshape, spin-lattice relaxation rate $1/T_{1}$
and spin-spin relaxation rate $1/T_{2}$ clearly indicates that they
are sensitive to the LRO onset and are driven by the fluctuations
of the magnetic V$^{4+}$ ions. In the vicinity of LRO, the existence
of two components is observed in the spin-spin relaxation data indicating
the co-existence of  non-LRO and magnetically ordered regions.

\section{Sample preparation, crystal structure, and experimental details}

\begin{figure*}
\centering{}\includegraphics[scale=0.55]{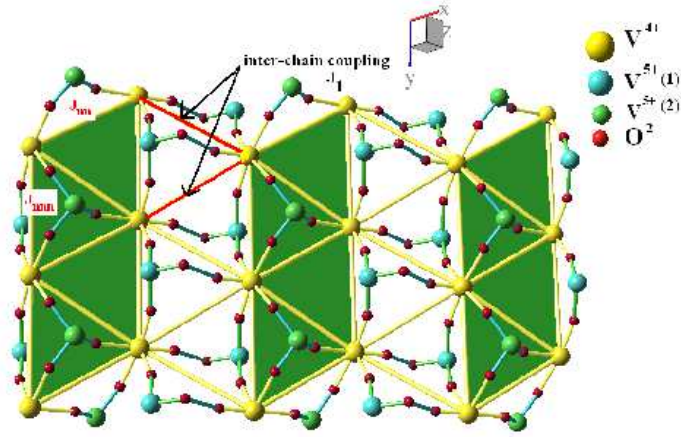}\caption{\label{fig:1}Schematic diagram of the coupled-MG chain network formed
by the V$^{\text{4+}}$ions. Possible interaction paths between the
V$^{\text{4+}}$ions are shown in the figure. }
\end{figure*}

BaV$_{\text{3}}$O$_{\text{8}}$ is a monoclinic system (space group
\textit{$P2_{\text{1}}/m$}) \cite{MarshJournal of solid state chemistry-BaV3O8}
containing both magnetic and non-magnetic vanadium ions. Among the
three vanadium ions in the unit cell, two are in the V$^{\text{5+}}$($S=0$,
non-magnetic) state and one is in the V$^{\text{4+}}$ ($S=1/2$)
state. The V$^{\text{4+}}$ ions appear to form chains where the interactions
between the magnetic vanadium ions are likely mediated through oxygen
(O$^{\text{2-}}$) and the non-magnetic V$^{5+}$ ions(see Fig.\ref{fig:1}).
Based on the structure, one expects comparable $nn$ and $nnn$ interactions
within the chain. The inter-chain interactions are expected to be
weaker. One might, therefore, think of this system as made of coupled
MG chains. Since no magnetic data have been reported on BaV$_{\text{3}}$O$_{\text{8}}$
as yet, we have pursued the problem further and report here the preparation,
magnetic susceptibility $\chi(T)$, heat capacity $C_{\mathrm{P}}(T)$,
and $^{51}$V NMR measurements on BaV$_{3}$O$_{8}$. 

\begin{figure}
\begin{centering}
\includegraphics[scale=0.36]{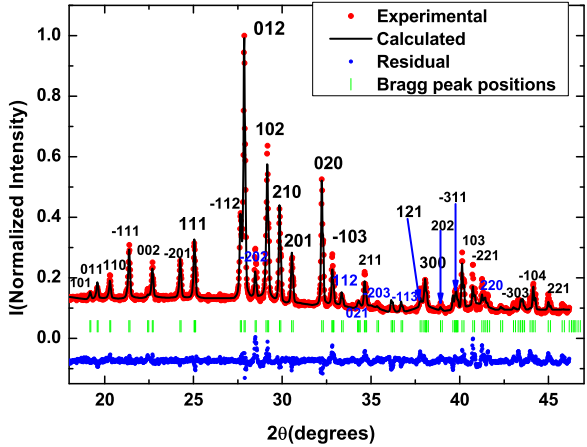}
\par\end{centering}

\caption{\label{fig:2}The x-ray diffraction pattern of BaV$_{\text{3}}$O$_{\text{8}}$
is shown along with its Bragg peak positions. The red points are the
experimental data, the black curve is the ``Fullprof'' generated
refinement pattern, the green markers are the Bragg peak positions
and the blue points represent the difference between the measured
and the calculated intensities.}
\end{figure}

BaV$_{\text{3}}$O$_{\text{8}}$ was prepared by standard solid state
reaction methods. In the first step we prepared BaV$_{\text{2}}$O$_{\text{6}}$
by firing stoichiometric amounts of BaCO$_{\text{3}}$ (Alfa Aesar-$99.95$\%)
and V$_{\text{2}}$O$_{\text{5}}$ (Aldrich-$99.6$\%) at $680$\textdegree{}C
for $28$ hours. In the next step we fired BaV$_{\text{2}}$O$_{\text{6}}$
and VO$_{\text{2}}$ (Alfa Aesar-$99.5$\%) at $700$\textdegree{}
for $36$ hours in a sealed quartz tube. X-ray diffraction (xrd) patterns
were collected with a PANalytical x-ray diffractometer using Cu $K_{\alpha}$
radiation ($\lambda=1.54182\textrm{\ensuremath{\mathring{A}}}$).
The Rietveld refinement of the xrd data was carried out using the
``Fullprof'' software\cite{Fullprof}( see Fig.\ref{fig:2}). Whereas
the refinement was attempted using two different monoclinic space
groups (\textit{$P2_{\text{1}}/m$}\cite{MarshJournal of solid state chemistry-BaV3O8}
and\textit{ $P2_{\text{1}}$}\cite{BaV3O8 space group P21}), we obtained
better results using the first one. The refined atomic coordinates
are given in table \ref{atomic positions} . 

\begin{table}
\begin{tabular}{|c|c|c|c|c|}
\hline 
Atoms & \multicolumn{1}{c}{} & \multicolumn{1}{c}{Coordinates} &  & Occupancy\tabularnewline
\hline 
 & x($\textrm{\ensuremath{\mathring{A}}}$) & y($\textrm{\ensuremath{\mathring{A}}}$) & x($\textrm{\ensuremath{\mathring{A}}}$) & \tabularnewline
\hline 
Ba(2e) & 0.212 & 0.250 & -0.099 & 1.00\tabularnewline
\hline 
V$^{\text{5+}}$(2e) & 0.072 & 0.250 & 0.361 & 1.00\tabularnewline
\hline 
V$^{\text{4+}}$(2e) & 0.294 & 0.750 & 0.216 & 1.00\tabularnewline
\hline 
V$^{\text{5+}}$(2e) & 0.585 & 0.250 & 0.316 & 1.00\tabularnewline
\hline 
O1(4a) & 0.480 & 0.494 & 0.200 & 1.00\tabularnewline
\hline 
O2(2e) & 0.432 & 0.750 & 0.476 & 1.00\tabularnewline
\hline 
O3(4a) & 0.156 & 0.505 & 0.261 & 1.00\tabularnewline
\hline 
O4(2e) & 0.167 & 0.750 & 0.019 & 1.00\tabularnewline
\hline 
O5(2e) & -0.171 & 0.250 & 0.319 & 1.00\tabularnewline
\hline 
O6(2e) & -0.145 & 0.250 & 0.545 & 1.00\tabularnewline
\hline 
\end{tabular}\caption{\label{atomic positions}Atomic positions in BaV\textsubscript{3}O\textsubscript{8}}

\end{table}

The cell parameters obtained for BaV$_{\text{3}}$O$_{\text{8}}$
(Space group: \textit{$P2_{\text{1}}/m$}, $a=7.432\textrm{\ensuremath{\mathrm{\AA}}}$,
$b=5.549\textrm{\ensuremath{\mathrm{\AA}}}$, $c=8.200\mathrm{\mathring{A}}$,
$\beta$= $107.207\text{\textdegree}$) are in agreement with those
of Ref.\cite{MarshJournal of solid state chemistry-BaV3O8}. The goodness
of the Rietveld refinement as defined by the following parameters
is R$_{\text{p}}$= $18.6$\%, R$_{\text{wp}}$= $14$\%, R$_{\text{exp}}$=$11.10$\%,
and $\lyxmathsym{\textgreek{q}}^{\text{2}}$ = $1.58$. By simple
valence counting, from among the three vanadium ions in the unit cell,
two would be in the V$^{\text{5+}}$ ($S=0$, non-magnetic) state
and one in the V$^{\text{4+}}$ ($S=1/2$) state. In the unit cell
of BaV$_{\text{3}}$O$_{\text{8}}$, V$^{\text{5+}}$ ions form tetrahedra
with the oxygen ions whereas the V$^{\text{4+}}$ ions form pyramids
with the oxygen ions. 

The various bond angles and bond lengths are given in Table \ref{Bond-angles}
and Table \ref{Bond lengths}. The possible interaction paths between
the V$^{4+}$ ions are illustrated in Fig.\ref{fig:1}. The V$^{\text{4+}}$
ions appear to be arranged in a coupled MG chain-like fashion (along
the crystallographic $b$-direction) and further form corrugated planes
as shown in Fig.\ref{fig:1}. If we concentrate on an MG chain it
can be seen that the V$^{\text{4+}}$ ions form triangular plaquettes
and in a given triangular plaquette the bond distances and the bond
angles between any two vanadiums are nearly the same. This suggests
a frustrated scenario of an MG chain with comparable $J_{nnn}$ and
$J_{nn}$. The coupling between the V$^{4+}$ ions of adjacent chains
($J_{1}$) (in a corrugated plane) might be weaker than the intra-chain
V$^{4+}$- V$^{4+}$ interaction since the inter-chain V$^{4+}$-
V$^{4+}$interaction path consists of one extra oxygen ion. There
could also be a weaker V$^{4+}$- V$^{4+}$interaction ($J_{2}$)
in a direction perpendicular to the corrugated plane of the coupled-MG
chains. Therefore, based on the structural details alone one might
expect SRO in this system. 

\begin{table}
\begin{tabular}{|c|c|c|}
\hline 
Angles & Description & Value\tabularnewline
\hline 
\hline 
V$^{\text{4+}}$-O$^{\text{2-}}$(1)-V$^{\text{5+}}$(2) & intra-chain & 137.07\textdegree{}\tabularnewline
\hline 
O$^{\text{2-}}$(1)-V$^{\text{5+}}$(2)-O$^{\text{2-}}$(2) & intra-chain & 113.08\textdegree{}\tabularnewline
\hline 
V$^{\text{5+}}$(2)-O$^{\text{2-}}$(2)-V$^{\text{4+}}$ & intra-chain & 147.75\tabularnewline
\hline 
V$^{\text{4+}}$-O$^{\text{2-}}$(3)-V$^{\text{5+}}$(1) & intra-chain & 163.76\textdegree{}\tabularnewline
\hline 
O$^{\text{2-}}$(3)-V$^{\text{5+}}$(1)-O$^{\text{2-}}$(3) & intra-chain & 101.01\textdegree{}\tabularnewline
\hline 
O$^{\text{2-}}$(1)-V$^{\text{5+}}$(2)-O$^{\text{2-}}$(1) & intra-chain & 104.89\textdegree{}\tabularnewline
\hline 
O$^{\text{2-}}$(3)-V$^{\text{5+}}$(1)-O$^{\text{2-}}$(5) & inter-chain & 112.44\textdegree{}\tabularnewline
\hline 
V$^{\text{5+}}$-(1)O$^{\text{2-}}$(5)-O$^{\text{2-}}$(4) & inter-chain & 83.15\textdegree{}\tabularnewline
\hline 
O$^{\text{2-}}$(5)-O$^{\text{2-}}$(4)-V$^{\text{4+}}$ & inter-chain & 145.43\textdegree{}\tabularnewline
\hline 
\end{tabular}

\caption{\label{Bond-angles} Bond angles between various vanadium and oxygen
ions in BaV\textsubscript{3}O\textsubscript{8}}
\end{table}

\begin{table}
\begin{tabular}{|c|c|c|}
\hline 
Bonds & Description & Value\tabularnewline
\hline 
\hline 
V$^{\text{4+}}$-O$^{\text{2-}}$(1) & intra-chain & 2.01Å\tabularnewline
\hline 
O$^{\text{2-}}$(1)-V$^{\text{5+}}$(2) & intra-chain & 1.71Å\tabularnewline
\hline 
V$^{\text{5+}}$(2)-O$^{\text{2-}}$(2) & intra-chain & 1.74Å\tabularnewline
\hline 
O$^{\text{2-}}$(2)-V$^{\text{4+}}$ & intra-chain & 2.07Å\tabularnewline
\hline 
V$^{\text{4+}}$-O$^{\text{2-}}$(3) & intra-chain & 1.81Å\tabularnewline
\hline 
O$^{\text{2-}}$(3)-V$^{\text{5+}}$(1) & intra-chain & 1.83Å\tabularnewline
\hline 
V$^{\text{5+}}$(1)-O$^{\text{2-}}$(5) & inter-chain & 1.74Å\tabularnewline
\hline 
O$^{\text{2-}}$(5)-O$^{\text{2-}}$(4) & inter-chain & 2.78Å\tabularnewline
\hline 
O$^{\text{2-}}$(4)-V$^{\text{4+}}$ & inter-chain & 1.62Å\tabularnewline
\hline 
\end{tabular}\caption{\label{Bond lengths}The bond lengths of the various vanadium-oxygen
linkages in BaV\textsubscript{3}O\textsubscript{8}}

\end{table}

\begin{figure}

\includegraphics[scale=0.5]{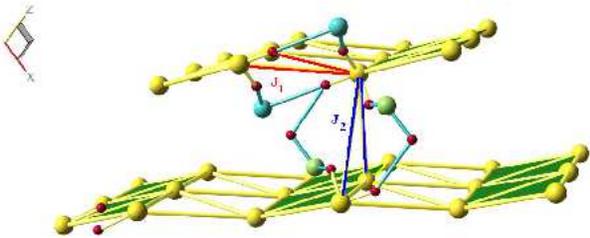}\caption{\label{fig:3}Possible inter-chain interaction paths between the V$^{\text{4+}}$ions }
\end{figure}

The temperature dependence of the magnetization $M$ was measured
in a magnetic field of 5 kOe in the temperature $T$ range $2-300$K
using a vibrating sample magnetometer (VSM) attached to a Quantum
Design Physical Property Measurement System (PPMS). The temperature
dependence of the specific heat has been measured in the temperature
range of $2-270$K by using a Quantum Design PPMS.The field-sweep
NMR spectra were recorded by means of a home-built phase coherent
pulsed spectrometer by sweeping the magnetic field at several fixed
frequencies. Typical pulse lengths were $5$ and $10$ \textgreek{m}s
for the $\lyxmathsym{\textgreek{p}}/2$ and $\lyxmathsym{\textgreek{p}}$
pulses, respectively, with a pulse separation of $50$ $\lyxmathsym{\textgreek{m}}$s.
At each temperature, the area under the spin-echo magnitude was integrated
in the time domain and averaged over mutiple scans. The \textsuperscript{51}V
nuclear spin-lattice relaxation was measured by the inversion recovery
method using a $\lyxmathsym{\textgreek{p}}$ - $t$ - $\lyxmathsym{\textgreek{p}}/2$
- $\tau$ - $\lyxmathsym{\textgreek{p}}$ pulse sequence with $\tau$
= $50$ $\lyxmathsym{\textgreek{m}}s$ and variable $t$. To obtain
the saturation magnetization $M_{0}$, the first \textgreek{p}-pulse
was switched off every even scan. Subtracting results of odd and even
scans one obtains $M_{t}$ which is generally fit to $M_{t}=-M_{0}+2M_{0}$$\times$
$e$$^{(-t/T_{1})}$. In the present case, two components were found
in the relaxation behaviour. Consequently, the above equation was
modified to accommodate two exponentials with different coefficients.
The spin-spin relaxation rate $1/T_{2}$ measurements were performed
in the temperature range of $1.65\lyxmathsym{\textendash}9.2$K at
a fixed frequency of $70$MHz on the maximum of the spectra (at an
applied field $62.6$ kOe). The decay of the spin echo (integrated
intensity) following a $\pi$/$2$ -$\pi$ pulse sequence was monitored
as a function the pulse separation $\tau$.

\section{Results}

\begin{figure}
\centering{}\includegraphics[scale=0.3]{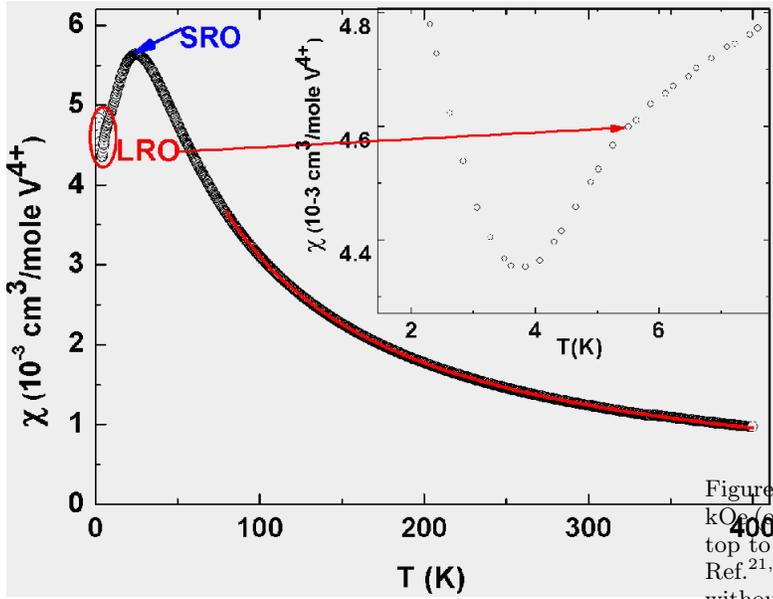}\caption{\label{fig:4}The temperature dependence of the susceptibility $\chi=M/H$
at $H=5$ kOe. LRO and SRO regions are indicated. The inset shows
a zoom of the low-$T$ region where the LRO region is marked. The
red line shows the Curie-Weiss fit in the temperature range of 80-300K. }
\end{figure}

With decreasing temperature, $\chi$ follows the Curie-Weiss law and
shows a broad maximum at about $25$K (see Fig.\ref{fig:4}). With
a further decrease in $T$, a sharp drop is observed in $\chi(T)$
at $T_{\mathrm{N}}\simeq6$K. A Curie-like upturn is observed at lower
temperatures. From the Curie-Weiss fit $\chi(T)$ = $\lyxmathsym{\textgreek{q}}$$_{\text{0}}$+
$C/(T-\theta_{\text{CW}})$ in the range $80-300$K, we get the $T$-independent
susceptibility $\mbox{\textgreek{q}}_{\text{0}}$ = $5.07\times10^{\text{-5}}$cm$^{\text{3}}$/mole
V$^{\text{4+}}$, the Curie constant $C=0.39$ cm$^{\text{3}}$K/mole
V$^{\text{4+}}$, and the Curie-Weiss temperature $\theta_{\text{CW}}=-26\mathrm{K}$.
With $S=1/2$ for the V$^{4+}$ ion, this Curie constant yields a
$g$-factor of $2.04$ indicating a very small spin-orbit coupling.
From $\mbox{\textgreek{q}}_{\text{0}}=5.07\times10^{\text{-5}}$cm$^{\text{3}}$/mole-V$^{\text{4+}}$we
obtain the Van Vleck susceptibility $\mbox{\ensuremath{\chi}}_{\text{VV}}=\mbox{\textgreek{q}}_{\text{0}}-\chi_{\mathrm{core}}=1.94\times10^{\text{-4}}$
cm$^{\text{3}}$/mole V$^{\text{4+}}$where $\chi_{\text{core}}$
is the core diamagnetic susceptibility equal to $-1.43\times10^{-4}$
cm$^{\text{3}}$/ mole formula unit.\cite{Core susceptibility} The
broad maximum at 25K might signify the onset of SRO in the MG chains.
The second anomaly observed at $T_{\mathrm{N}}$ (see inset of Fig.\ref{fig:4})
perhaps evidences the onset of LRO. With the frustration parameter
$f=\frac{\left|\theta_{\mathrm{CW}}\right|}{T_{\mathrm{N}}}\sim5$
it appears that the system is moderately frustrated.\cite{Shahab f>5} 

\begin{figure}

\includegraphics[scale=0.3]{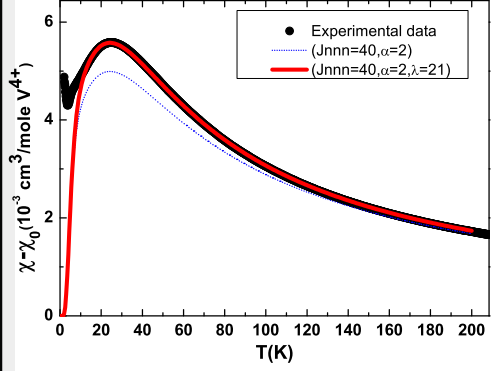}\caption{\label{fig:5}The temperature dependence of $\chi-\chi_{0}$ at $H=5$
kOe (open circles). The red line and the blue dotted line (from top
to bottom) represent the exact diagonalization results of Ref. \cite{MG-chain-simulation,MG chain-simulated data}
with inter-chain interaction ($\lyxmathsym{\textgreek{a}}$ = 2, $\lambda=21$)
and without inter-chain interaction ($\lyxmathsym{\textgreek{a}}$
= $2$, $\lambda=0$), respectively.}

\end{figure}

The magnetic susceptibility of BaV$_{3}$O$_{8}$ was modeled using
the exact diagonalization results for the susceptibility $\lyxmathsym{\textgreek{q}}_{\mathrm{MG}}$($g$,
$\lyxmathsym{\textgreek{a}}$, $J_{nn}$) of a single chain provided
by Heidrich-Meissner et al.\cite{MG-chain-simulation,MG chain-simulated data}
(where $\lyxmathsym{\textgreek{a}}=J_{nnn}/J_{nn}$). Since our experimental
results did not match those of the isolated Majumdar-Ghosh chain simulation
for different parameters $\alpha$ and $J_{nn}$, we used the expression
for a coupled Majumdar-Ghosh chain model, where the inter-chain spin
exchange is treated within a mean-field approach, \cite{mean-field- reference}
as eqn.\ref{eq:coupledMG}

\begin{equation}
\lyxmathsym{\textgreek{q}}_{\mathrm{coupled\, MG}}(T)=(\lyxmathsym{\textgreek{q}}_{\mathrm{MG}}/(1-\lambda\lyxmathsym{\textgreek{q}}_{\mathrm{MG}}))+\lyxmathsym{\textgreek{q}}_{\text{0}}\label{eq:coupledMG}
\end{equation}
Here, $\lambda$ is the mean-field parameter

\begin{equation}
\lambda=(z_{\text{1}}J_{1}+z_{\text{2}}J_{2})/N_{A}g^{2}\mu_{B}^{2}\label{eq:meanfieldparameter}
\end{equation}
where $J_{1}$ and $J_{2}$ are the inter-chain coupling constants
within the plane and perpendicular to the plane, respectively, as
shown schematically in Fig.\ref{fig:1} and Fig.\ref{fig:3}. The
corresponding number of neighbours is $z_{1}$ and $z_{2}$. We found
the best agreement of our data with the coupled MG chain model calculations
for $\lyxmathsym{\textgreek{a}}$ = $2$, $J_{nnn}$/$k_{B}$ = 40K
, and $\lambda=21$mol/cm\textsuperscript{3}(cf. red solid line in
Fig. \ref{fig:5}). Taking $z_{1}=z_{2}=2,$(see Fig.\ref{fig:1}
and Fig.\ref{fig:3}) we get $J_{1}+J_{2}=16$K. We remark in passing
that the MG point itself is exemplified by $\lyxmathsym{\textgreek{a}}$
= $0.5$. The positive sign of the net inter-chain coupling constant
indicates a ferromagnetic interaction. The significant inter-chain
interactions seem to drive the system to a LRO state at low temperature.
Whereas fitting of our bulk susceptibility data are indeed suggestive
of a coupled MG chain scenario, we caution that any definitive conclusions
need to be backed up by additional work such as neutron scattering
measurements, density functional theory calculations to estimate the
relative exchange couplings.

\begin{figure}
\centering{}\includegraphics[scale=0.35]{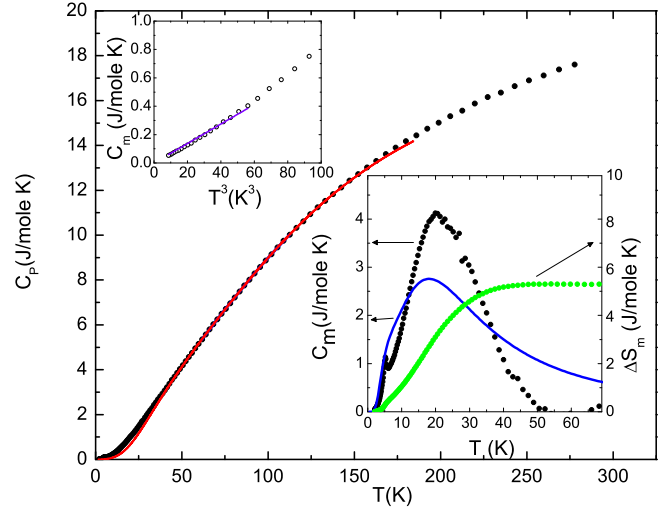}\caption{\label{fig:6}The temperature dependence of specific heat of BaV$_{\text{3}}$O$_{\text{8}}$;
the inset on the right displays the magnetic contribution to the heat
capacity (black filled circles). The red line represents the lattice
heat capacity (see text). The blue line in the lower inset represents
the exact diagonalization results of Ref. \cite{MG-chain-simulation,MG chain-simulated data}
with ($\lyxmathsym{\textgreek{a}}$ = $2$, $\lambda=0$). Note that
the total heat capacity of the main figure is normalised to the number
of atoms ($12$ per formula unit) whereas the magnetic heat capacity
in the inset is normalised to the number of magnetic atoms, $i.e.$,
one for each formula unit. The green data points (right axis) show
the change of entropy with $T$. The upper inset shows the linear
temperature dependence of the magnetic specific heat with $T$\textsuperscript{3}.
The solid line is a linear fit to low temperature data.}
\end{figure}

The heat capacity data of BaV$_{3}$O$_{8}$ are shown in Fig.\ref{fig:6}
and exhibit a sharp anomaly at about $5.8$K. This is close to the
transition temperature $T_{\mathrm{N}}$ measured in $\chi(T)$. The
transition temperature does not shift with $H$ up to $90$ kOe. We
have modeled the lattice contribution using the Debye model in the
$T$-range $60-110$K. The measured heat capacity could be fit with
a combination of two Debye functions of the type given below with
coefficients $C_{1}$ and $C_{2}$(see Fig.\ref{fig:6}),

\begin{flushleft}
\begin{equation}
\mbox{\ensuremath{\mbox{\ensuremath{\mbox{C}_{p}}}}=\ensuremath{9Nk_{\text{B}}n}\ensuremath{(T/\theta_{\text{D}})^{\mbox{\text{3}}}}\ensuremath{\int_{\text{0}}^{\text{\ensuremath{\theta\ensuremath{_{\text{D}}}/T}}}(x^{\text{4}}e^{\text{\ensuremath{x}}}/(e^{\text{\ensuremath{x}}}-1)^{\text{2}}}\ensuremath{dx}}
\end{equation}
where $n$ is the number of atoms in a formula unit, $N$ is the Avogadro
number, $k_{\text{B}}$ is the Boltzmann constant, and $\theta_{\text{D}}$
is the relevant Debye temperature. The fit yields Debye temperatures
of $180$K and $580$K and their coefficients $0.25$ and $0.52$,
respectively. Upon subtracting the lattice heat capacity with the
above parameters, we obtain the magnetic contribution to the heat
capacity $C_{\mathrm{m}}(T)$. 
\par\end{flushleft}

Subsequently, the entropy change ($\Delta S)$ was calculated by integrating
the $C_{\text{m}}/T$ data (see Fig.\ref{fig:6}, right inset). The
entropy change from about $50$K to $2$K is about $5.4$ J/K which
is more than $90$\% of the value for a $S=1/2$ system ($Rln(2S+1))$.
The small disparity may be associated with the uncertainty in extracting
the lattice contribution. Note also that, upon cooling, most of the
entropy decrease has already taken place above $T_{\mathrm{N}}$.
This is a fingerprint of strong intra-chain interactions in the system
which give rise to SRO. The heat capacity of a model MG chain corresponding
to $\alpha=2$ and $J_{nnn}/k_{B}=40$K is shown in the Fig.\ref{fig:6}
(right inset) by a blue solid line. The experimental magnetic heat
capacity is similar in magnitude to the calculated curve. The small
difference seen here can easily arise from uncertainties associated
with the estimation of the lattice heat capacity. 

We observed a linear dependence of magnetic heat capacity with $T$\textsuperscript{3}
below 4K (see Fig.\ref{fig:6}, left inset) which suggests the presence
of antiferromagnetic magnons in the ordered state\cite{magnon-antiferromagnet-magnetic sp. heat}. 

\begin{figure}
\begin{centering}
\includegraphics[scale=0.38]{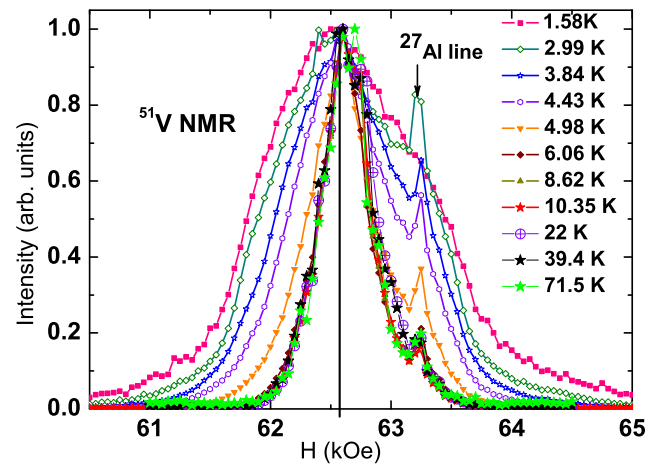}
\par\end{centering}

\centering{}\caption{\label{fig:7}The $^{51}$V spectra of BaV$_{\text{3}}$O$_{\text{8}}$
at different temperatures obtained by sweeping the field (the $^{27}$Al
signal is extrinsic and originates from the probehead) at a frequency
of $70$MHz. }
\end{figure}

\begin{figure}
\includegraphics[scale=0.33]{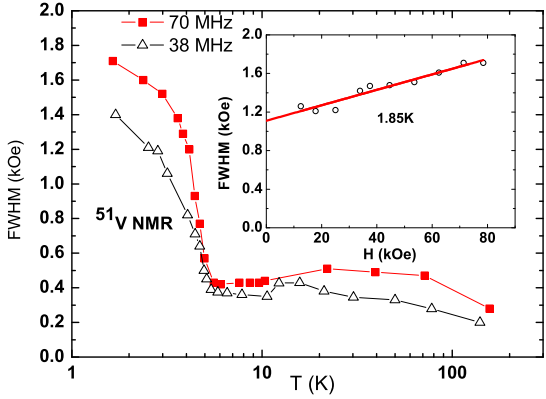}

\caption{\label{fig:8}Temperature dependence of FWHM measured at fixed frequencies
of $38$MHz (black open triangles) and $70$MHz (red closed squares)
for BaV$_{\text{3}}$O$_{\text{8}}$. The pulse separation is $50\mu s$
between the pulses of the spin-echo sequence irradiating $^{51}$V
nuclei. The inset shows the small variation in FWHM with the change
in magnetic field measured at $1.85$K. The red line is a linear fit
indicating the slight decrease in FWHM with decrease in magnetic field.}

\end{figure}

We were unable to detect the NMR signal associated with the magnetic
V$^{\text{4+}}$nuclei of BaV$_{\text{3}}$O$_{\text{8}}$. This is
because of a strong on-site local moment which naturally couples well
with its own nucleus. The fluctuations of this moment are very effective
in causing a fast relaxation of the nuclear magnetization. This makes
the detection of its NMR signal difficult. In Cs$_{\text{2}}$CuCl$_{\text{4}}$
as well, an NMR signal from the $^{63,65}$Cu nuclei was not detected
probably for similar reasons. \cite{M.A.Vachon New journal of physics-8(2006)-133Cs-NMR-Cs2CuCL4}
On the other hand, the NMR signal from the nonmagnetic V\textsuperscript{5+}
nuclei in BaV\textsubscript{3}O\textsubscript{8} was easily observed.
The spectrum is rather broad at $71.5$K with the full width at half
maximum FWHM of about $0.43$ kOe and the total extent of the spectrum
of about $1.6$ kOe (see Fig.\ref{fig:7}). The FWHM increases slightly
with decreasing temperature down to about $6$K and increases drastically
below that (see Fig.\ref{fig:8}) which must be associated with the
onset of LRO already indicated in $\chi(T)$ and $C_{\mathrm{P}}(T)$
data. This implies that the V$^{5+}$ nuclei are sensitive to the
internal magnetic field arising in this compound. It is shown in Fig.\ref{fig:8}
that FWHM in LRO phase ($T=1.85K$) decreases with $H$ although it
tends to the value of about $1.1$ kOe in zero field which is $2-3$
times more than the FWHM above $T_{\mathrm{N}}$. Thus the externally
applied magnetic field $H$ is not the only source of such a broadening
however the $^{51}$V NMR shift does not change much with temperature
which indicates that the V$^{\text{5+}}$ions are very weakly coupled
with the electrons of the magnetic vanadium (V$^{\text{4+}}$) ions. 

\begin{figure}
\centering{}\includegraphics[scale=0.33]{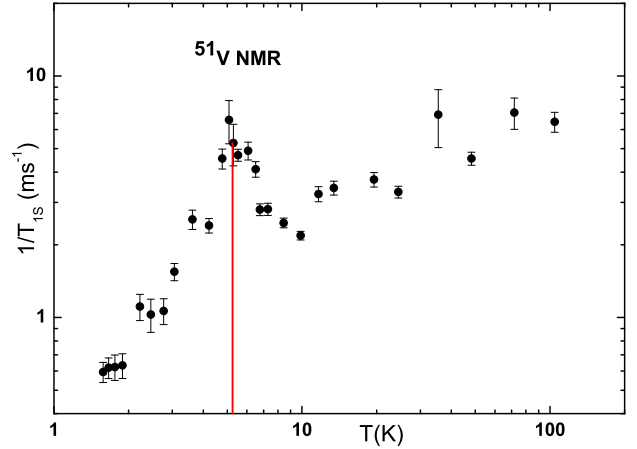}\caption{\label{fig:9}The temperature dependence of the spin-lattice relaxation
rate (1/$T_{1}$) corresponding to the faster component. The red vertical
line marks $T_{\mathrm{N}}$.}
 
\end{figure}

We have measured the recovery of the longitudinal $^{51}$V nuclear
magnetization as a function of temperature to probe the low-energy
spin excitations. The resulting recovery could be fitted well with
a double exponential (consisting of a short and a long component )
at all temperatures. The recovery was found to follow

\begin{equation}
1-M(t)/M_{o}=B_{s}e^{(-t/T_{1S})}+B_{L}e^{(-t/T_{1L})}+C_{1}
\end{equation}
Here $T_{1S}$ and $T_{1L}$ represent the short and the long components
of $T_{1}$, $B_{S}$ and $B_{L}$ stand for their relative weights
respectively and C\textsubscript{1} is a constant. \cite{Footnote-T1 I 7/2-1}
As seen in the structure, two inequivalent V\textsuperscript{5+}
ions are present in BaV\textsubscript{3}O\textsubscript{8}(Fig.\ref{fig:1}).
One of them (V$^{5+}$(2)) is near the centre of a triangular plaquette
and appears coupled to three V\textsuperscript{4+} ions. The other
vanadium (V$^{5+}$(1)) seems coupled to two V\textsuperscript{4+}
ions via oxygen. Whereas it is not clear as to the relative magnitudes
of the associated transferred hyperfine couplings in the above two
cases, the relevant bond angles and bond lengths suggest that the
various V$^{4+}$-O-V$^{5+}$ couplings may not be too much different
from each other. Therefore, the V\textsuperscript{5+}(2) which is
hyperfine coupled to three V$^{4+}$ might be expected to have a shorter
T$_{1}$ compared to that for V$^{5+}$(1). The variation of the faster
rate $1/T_{1S}$ with temperature is shown in Fig.\ref{fig:9} (the
slower component has the same qualitative behaviour). As seen from
this figure, a distinct anomaly of $1/T_{1}$ is observed at $T_{\mathrm{N}}$. 

\begin{figure}
\includegraphics[scale=0.33]{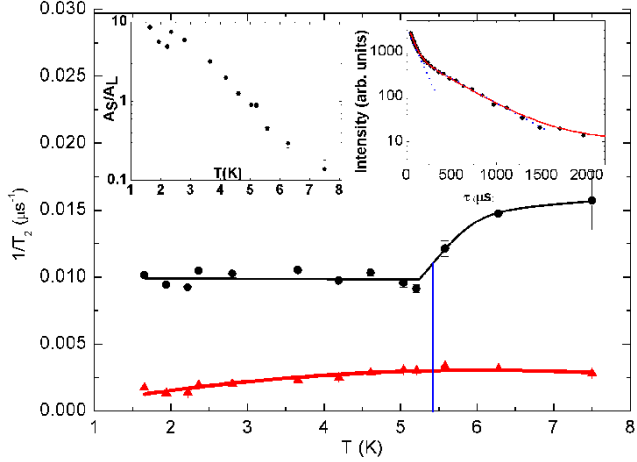}\caption{\label{Fig.10}The variation of the $^{51}V$ spin-spin relaxation
rates $1/T_{2S}$ and $1/T_{2L}$ with temperature is shown. The solid
lines are drawn as guides to the eye. The blue vertical line points
to the LRO temperature. The left inset shows the variation of the
relative weight of the fast to slow component of $T_{2}$ with temperature.
The right inset shows the spin-spin relaxation curve at $2.22$K.
Here, the black circles correspond to experimental data and the red
curve is a double exponential fit. The blue dashed lines correspond
to the fast relaxing and slow relaxing components.}
\end{figure}

Additionally, we measured the temperature dependence of the transverse
decay and obtained the spin-spin relaxation rates $1/T_{2}$ presented
in figure \ref{Fig.10}. As seen from the raw data at $2.22$K in
Fig.\ref{Fig.10}(right inset), the decay has two components shown
by the two dashed lines. Consequently, we have fit the data at each
temperature to a double exponential function

\begin{equation}
M_{t}=A_{s}e^{(-2\tau/T_{2S})}+A_{L}e^{(-2\tau/T_{2L})}+C_{2}
\end{equation}
Here, $T_{2S}$ and $T_{2L}$ denote the shorter and the longer components,
respectively, $A_{S}$ and $A_{L}$ stands for their relative weights
respectively and C\textsubscript{2} is a constant. The variation
of $1/T_{2S}$ and $1/T_{2L}$ with temperature is shown in Fig.\ref{Fig.10}.
As seen from Fig. \ref{Fig.10}, the short component, $1/T_{2S}$,
exhibits a pronounced $\sim$$50$\% decrease in the vicinity of $T_{\mathrm{N}}$
while the longer one, $1/T_{2L}$, is insensitive to the magnetic
ordering. The relative weight of the faster component ($A_{S}/A_{L}$)
decreases monotonically with increasing temperature (as seen in the
inset of Fig.\ref{Fig.10}) varying from about $10$ at $1.5$K to
about $0.1$ at about $7.5$K. As is evident, at higher temperatures
the spin-spin relaxation is dominated by the longer component. There
is, therefore, a larger uncertainty in $T_{2S}$ at higher temperatures.
We also note that there is no sharp anomaly in the temperature dependence
of $A_{S}/A_{L}$ near $T_{\mathrm{N}}$. Finally, it seems natural
to think that $A_{S}/A_{L}$represents the relative amount of magnetically
ordered regions with respect to non-LRO regions. 

\begin{figure}
\includegraphics[scale=0.36]{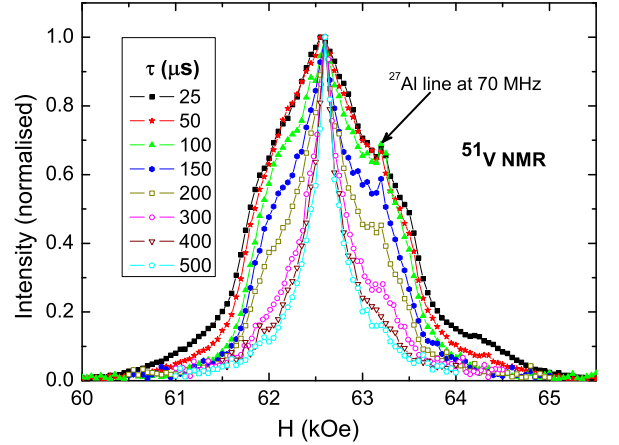}\caption{\label{Fig.11}The variation of $^{51}$V NMR spectra ($\nu=70$MHz
and $T=2.35$K) with the separation of pulses $\tau$. }
\end{figure}

\begin{figure}
\includegraphics[scale=0.33]{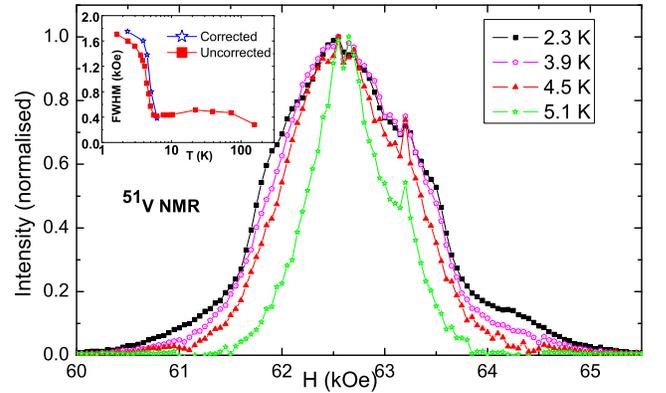}

\caption{\label{Fig.12}The corrected $^{51}$V NMR spectra containing only
the short $T_{2}$ component are shown at different temperatures (see
text for details). Line width vs. temperature plot for these spectra
are shown in the inset where the filled symbols represent uncorrected
data while the open symbols are identical data from Fig. \ref{fig:8}
at $70$MHz. }
\end{figure}

Given the above information, it is clear that the $^{51}$V NMR lineshape
in the low-temperature regime is affected by the measurement conditions,
in particular, by the pulse separation $\tau$ between the $\pi/2$
and the $\pi$ pulse. This is clearly seen in Fig.\ref{Fig.11} where
the $^{51}$V spectra powder pattern is displayed at fixed frequency
and temperature, but for different pulse separations $\tau$. A broader
resonance line is observed for smaller $\tau$. This might be the
reason for the observation of Fig.\ref{fig:8} where the FWHM keeps
increasing with decreasing temperature without saturation even well
below $T_{\mathrm{N}}$. It is also worth noting that even above $T_{\mathrm{N}}$,
where no ordered moments are expected, a contribution to the width
from the shorter $T_{2}$ remains. We have attempted to obtain the
contribution to the lineshape only from the faster $T_{2S}$ component
(i.e., the one originated from the magnetic regions) in the following
manner. With a large pulse separation like $\tau$ = $500$ $\mu s$,
the lineshape contains the contribution only from the slow component.
This is corrected using $T_{2L}$ and the spectrum from only the long
component is obtained at $\tau=0$. Then with our shortest pulse separation
$\tau=25$$\mu s$, the spectrum was measured. This contains both
the fast and the slow components with dominating fast component at
low temperatures. This is corrected for $T_{2S}$ and the spectrum
at $\tau=0$ is obtained. We then subtract the first data set ($T_{2L}$
dominated) from the latter. The resulting spectrum contains the contribution
of only the faster $T_{2}$ component (one can see though that in
the temperature region where $A_{S}/A_{L}\sim<1$, this method suffers
from a large uncertainty). The resulting spectra at various temperatures
are shown in Fig.\ref{Fig.12}. The inset shows the linewidth obtained
from such spectra which arises only from the magnetic regions. It
is higher than the uncorrected data and also has a sharper variation
near $T_{\mathrm{N}}$. 

\section{Conclusion}

In this work we have reported the crystal structure, magnetization,
heat capacity, and NMR measurements on a new vanadium-based magnetic
system BaV$_{\text{3}}$O$_{\text{8}}$. According to the crystallographic
symmetry, the arrangement of the magnetic ions in BaV$_{\text{3}}$O$_{\text{8}}$
is like that of a coupled Majumdar-Ghosh chain with the possibility
of three dimensional interactions. We found that the magnetic susceptibility
shows a broad maximum at about $25$K indicating magnetic short-range
order followed by a sharp anomaly at $T_{N}$ = $6$K due to long-range
order. The value of the \textquoteleft{}frustration parameter\textquoteright{}
($f=|\theta/|T_{\text{N}}|\sim5$) suggests that the system is moderately
frustrated. Our susceptibility data above 15K are well described by
the coupled Majumdar-Ghosh chain model with the ratio of the $nnn$
to $nn$ magnetic coupling $\lyxmathsym{\textgreek{a}}$ = $2$ and
$J_{nnn}$/$k_{B}$ = 40K. Considering the inter-chain interactions
in the mean-field approach, we obtain the total inter-chain coupling
$J_{nn}$/$k_{B}$ = $16$K. However, a validation of the above conclusions
would need input from techniques such as inelastic neutron scattering,
density functional theory calculations, etc. From the magnetic contribution
of the heat capacity $C_{\mathrm{m}}$ we find that upon cooling,
most of the entropy decrease has already taken place at temperatures
above $T_{\mathrm{N}}$. This attests to the strong intra-chain interactions
in the system which give rise to SRO. Below the LRO temperature of
$6$K, the magnetic heat capacity follows well a $T^{3}$ behaviour
suggesting antiferromagnetic magnon excitations. The value of the
entropy change ($\Delta S$) calculated from the magnetic specific
heat is nearly equal to that expected for a $S=1/2$ system. In the
local probe measurements (NMR) we were unable to detect the signal
originating from the magnetic vanadium due to the fast on-site fluctuations
of the V\textsuperscript{4+} local moments to which the \textsuperscript{51}V
nucleus is strongly coupled and consequently giving rise to very short
relaxation times. The $^{51}$V signal that was observed by us stems
from the non-magnetic vanadium sites. We found a clear signature of
the formation of long-range magnetic order below $T_{\mathrm{N}}$
from the temperature dependent NMR line width. Further, we found that
the spin-spin relaxation consists of two components. We suggest that
the shorter component arises from a coupling through the magnetic
regions while the longer one is from non-LRO regions. Below $T_{\mathrm{N}}$,
both non-magnetic and magnetically ordered regions are found to coexist
which, possibly, is characteristic of the inherent frustration in
the system. In view of a unique V$^{4+}$ site in the structure, the
coexistence might imply the presence of lattice distortions at low
temperatures creating inequivalent magnetic environments. Further
work using techniques such as muon spin rotation and neutron diffraction
is needed to explore this aspect.

\section{Acknowledgement}

Discussions with R.K.Sharma are acknowledged. T.C., A.V.M., A.A.G.,
and A.V.T. thank the Joint Indo-Russian RFFI-DST Grant 11-02-92707-
IND for financial support. W.K. and N.B. kindly acknowledge support
from the German Research Society (DFG) via TRR80 (Augsburg, Munich).

\end{document}